\documentclass[lettersize,journal]{IEEEtran}
\usepackage{xcolor}
\usepackage{amsmath,amsfonts}
\usepackage{algorithmic}
\usepackage{algorithm}
\usepackage{array}
\usepackage[caption=false,font=normalsize,labelfont=sf,textfont=sf]{subfig}
\usepackage{textcomp}
\usepackage{stfloats}
\usepackage{url}
\usepackage{verbatim}
\usepackage{graphicx}
\usepackage{cite}
\usepackage[numbers]{natbib}
\begin{document}

\title{A Constant-Gain Equation-Error Framework for Airliner Aerodynamic Monitoring Using QAR Data}

\author{Ruiying Wen, Yuntao Dai$^{*}$, Hongyong Wang
        % <-this % stops a space
\thanks{$^*$Yuntao Dai is the corresponding author.}
\thanks{Ruiying Wen is with the Civil Aviation University of China, Dongli District, Tianjin, China}% <-this % stops a space
\thanks{Yuntao Dai is with the Civil Aviation University of China, Dongli District, Tianjin, China}
\thanks{Hongyong Wang is with the Civil Aviation University of China, Dongli District, Tianjin, China}
\thanks{Manuscript received July 27, 2025; revised November 9, 2025; Accepted January 2, 2026.}}

\markboth{IEEE Transactions on Intelligent Transportation Systems,~Vol.~27, No.~5, May~2026}%
{Shell \MakeLowercase{\textit{et al.}}: A Sample Article Using IEEEtran.cls for IEEE Journals}

\IEEEpubid{1558-0016~\copyright~2026 IEEE. Personal use of this material is permitted. Permission from IEEE must be obtained for all other uses.}

\maketitle

\begin{abstract}
Monitoring in-service aerodynamic performance of airliners is critical for operational efficiency and safety, yet presents significant challenges when using operational Quick Access Recorder (QAR) data due to sensor noise, low excitation, and the absence of key model parameters like moments of inertia. These constraints render conventional state-propagation and recursive estimation methods unsuitable. To address these challenges, this paper proposes and validates the Constant-Gain Equation-Error Method (CG-EEM), a robust framework tailored for QAR data analysis. The CG-EEM employs a custom constant-gain estimator that avoids both the infeasibility of state-propagation filters and the premature convergence or instability issues of standard recursive algorithms. Extensive validation on a multi-fleet dataset of over 200 flights demonstrates that the framework produces highly consistent and physically meaningful aerodynamic parameters. Crucially, follow-up work has verified that this approach successfully resolves the fundamental thrust-drag ambiguity problem, ensuring the estimates are not just plausible, but physically unique and correct. Proving CG-EEM is a scalable and computationally efficient tool for reliable fleet-wide performance monitoring and early detection of airframe degradation.
\end{abstract}


\section{INTRODUCTION}
\label{introduction}

An aircraft's aerodynamic characteristics are not static; they evolve throughout its operational life. Factors such as structural aging, engine wear, and surface roughness from paint degradation or contamination can significantly alter performance from the original design specifications. Furthermore, modifications like the installation of new antennas can introduce unexpected aerodynamic penalties. Research has quantified these effects, showing that coating degradation can increase cruise drag by up to 7.1\% on a Boeing 737-NG, while a fuselage-mounted Wi-Fi antenna can add another 1.92\% \cite{does}. This "performance drift" directly impacts fuel efficiency and can alter handling qualities, underscoring the critical need for methods capable of continuously monitoring an aircraft's true aerodynamic state. Fortunately, modern airliners record a wealth of data using systems like the Quick Access Recorder (QAR). This presents a vast, yet largely untapped resource, for addressing this challenge, with emerging research exploring its use for applications like flight quality analysis \cite{wang2020data}.

Traditionally, Aerodynamic Parameter Estimation (APE) has relied on dedicated flight testing and computational fluid dynamics (CFD) \cite{APEapplication, systemidentification, CFD, CFDChallenge}. While these "gold-standard" approaches provide high-fidelity results, they are too resource-intensive for routine monitoring, creating a need for methods that can leverage in-service operational data.

Two modern paradigms have emerged to fill this gap. The first, purely data-driven modeling using neural networks, has shown promise in capturing complex aerodynamic relationships \cite{NNreduced, RNNreduced, LSTM-APE,DataDriven}. However, their "black-box" nature presents a major obstacle to certification and physical interpretation in safety-critical aviation \cite{XAI}. The second paradigm, a physics-informed filtering approach, has centered on joint-estimation filters, mainly Kalman filters and their mutations \cite{EKFUKF,Bay,EKFQAR,CGKF}. These methods attempt to simultaneously estimate both the aircraft state and its aerodynamic parameters. However, they rely on propagating the aircraft state through a dynamic model, which faces a fundamental barrier in a real-world operational context. Accurate state propagation, particularly of rotational dynamics, requires the aircraft's moments and products of inertia. These mass properties change continuously with fuel burn and are not recorded in standard QAR data, making a rigorous physics-based state propagation model impossible to implement. This unresolvable model error renders such coupled-filtering approaches fundamentally unsuitable for this specific application.

The limitations of state-propagation filters thus necessitate a decoupled, Equation-Error Method (EEM). In an EEM framework, the estimation problem is simplified by using the measured states from QAR data directly as inputs to an algebraic model of the aerodynamic forces, bypassing the need for dynamic propagation. The most common algorithm for such problems is the standard Recursive Least Squares (RLS) estimator.

\IEEEpubidadjcol
However, a critical analysis reveals that conventional RLS is also not suited for this application due to the nature of its time-varying gain. RLS is designed to progressively reduce its estimation gain as it processes more data, effectively giving more weight to initial samples. In the context of noisy and low-excitation cruise data, this behavior is detrimental. The estimator can prematurely converge on inaccurate parameters based on initial sensor noise, or become low sensitive before it has processed enough data to find the true underlying signal. While a forgetting factor can be used to keep the gain from vanishing, this often leads to instability, causing the parameter estimates to drift and chase noise rather than converge.

To overcome these dual challenges, this paper proposes and validates the Constant-Gain Equation-Error Method (CG-EEM). Our framework is built on the robust EEM principle to bypass the need for unavailable aircraft parameters. Crucially, it replaces the standard time-varying gain estimator with a custom-designed, constant-gain algorithm inspired by the Constant-Gain Kalman Filter (CGKF) \cite{CGKF}. This approach is perfectly suited to extracting precise parameter estimates from long segments of stationary, noisy data by ensuring every data point contributes consistently to the final result.

This framework provides a cost-effective and scalable tool for airlines to implement fleet-wide aerodynamic health monitoring. The identified parameters can be directly integrated into Flight Operations Quality Assurance (FOQA) programs for early anomaly detection, fed into fuel management systems to optimize flight plans with aircraft-specific performance models, and used to guide proactive maintenance scheduling, thereby enhancing both operational safety and fuel efficiency.

The primary contributions of this work are threefold:
\begin{enumerate}
    \item \textbf{The development of CG-EEM,} a robust framework for aerodynamic estimation designed specifically for the constraints and characteristics of operational QAR data.
    \item \textbf{A rigorous, simulation-based comparative analysis} demonstrating that the CG-EEM is superior to conventional time-varying gain RLS for this application, avoiding both premature convergence and noise-induced instability.
    \item \textbf{Extensive validation on a large, multi-fleet dataset} of over 200 flights, confirming the framework's consistency and scalability. Crucially, we demonstrate its ability to produce physically meaningful aerodynamic parameters, whose fidelity is independently verified in our follow-on work to overcome the inherent thrust-drag ambiguity problem, ensuring the physical consistency of the estimates. The details of this validation methodology will be presented in a subsequent paper.
\end{enumerate}

\section{QUICK ACCESS RECORDER (QAR) DATA}

\label{QAR}
Modern commercial airliners generate vast quantities of operational data through systems designed for proactive monitoring, complementing the traditional post-accident recorders (FDR and CVR). The primary source for this data is the Quick Access Recorder (QAR), which records thousands of flight parameters at high fidelity. The evolution of this technology to the Wireless QAR (WQAR) has automated post-flight data retrieval, while future systems may provide in-flight data transmission, further enhancing monitoring capabilities. For analysis, this data is decoded from its proprietary binary format and commonly stored in HDF5 files, which is ideal as it preserves not only the time-series measurements but also essential metadata, including sensor units and data types. The key parameters used in this study are summarized in Table~\ref{QARtable}.

\begin{figure}
    \centering
    \includegraphics[width=0.8\linewidth]{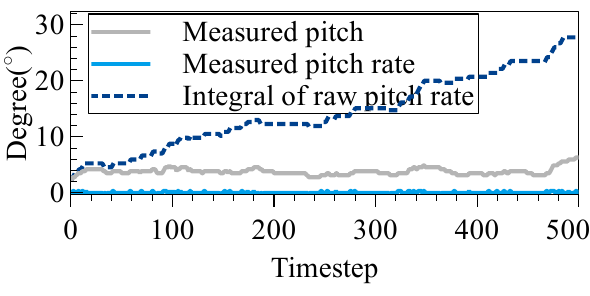}
    \caption{Comparison of integrated pitch rate and measured pitch angle, demonstrating significant deviation due to sensor noise.}
    \label{QARdeviation}
\end{figure}

Despite its richness, using raw QAR data for accurate Aerodynamic Parameter Estimation (APE) presents several fundamental challenges that preclude the direct application of standard system identification techniques:
\begin{enumerate}
    \item \textbf{Sensor Noise and Quantization:} Onboard sensors, while adequate for flight operations, have inherent precision and granularity limitations. Fig.~\ref{QARdeviation} provides a stark illustration of this problem, comparing the aircraft's measured pitch angle with the time-integral of the raw pitch rate signal. The significant drift and divergence of the integrated signal demonstrate how even small amounts of sensor noise are amplified through integration, leading to physically inconsistent results. Any robust APE framework must effectively avoid the detrimental effects of this measurement noise.
    \item \textbf{Lack of Critical Model Parameters:} Established APE frameworks, particularly those based on full non-linear equations of motion, often require parameters that are not explicitly recorded in QAR data. The most significant of these include manufacturer-specific thrust data and, crucially, the aircraft's moments and products of inertia. Without these, accurate state propagation for dynamic models is fundamentally infeasible, rendering coupled-estimation approaches (as discussed in Section~\ref{introduction}) unsuitable.
    \item \textbf{Low Persistent Excitation:} Operational flight data, particularly during quasi-steady, level cruise segments, lack the dynamic variation typically needed for many estimators to converge. Dedicated flight tests are designed with specific maneuvers (e.g., doublets, frequency sweeps) to deliberately excite dynamic modes. In contrast, the vast majority of operational data is recorded during trim conditions, providing limited information content per sample. Any framework intended for routine monitoring must therefore be robust enough to extract meaningful information from these less-than-ideal conditions.
\end{enumerate}

\begin{table*}
	\caption{Explanation and nomenclature of QAR data used in this paper. Computed data are in 1 Hz.}
	\centering{
		\begin{tabular}{c c c c c c}
			\hline
			 QAR Code   & Physical Meaning            & Unit       & Symbol     & Granularity & Sampling frequency \\ \hline
			    TAS     & True Air Speed              & knots      & $V$        & 0.0001      & Computed               \\
			    GW      & Gross Weight                & lbs        & $m$        & 0.0001  & 1/64 Hz               \\
			   VRTG     & Vertical g-acceleration     & g          & $a_{z}$   & 0.0039      & 16 Hz               \\
			   LONG     & Longitudinal g-acceleration & g          & $a_{x}$   & 0.0039      & 16 Hz               \\
			   PITCH    & Pitch Angle                 & $^\circ$   & $\theta$   & 0.0001      & 4 Hz               \\
			 FLT\_PATH  & Flight Path Angle           & $^\circ$   & $\gamma$   & 0.0001      & 1 Hz               \\
			PITCH\_RATE & Pitch Rate                  & $^\circ$/s & $q$        & 0.0001      & 8 Hz               \\
			    TAT     & Total Air Temperature       & $^\circ$C  & $T$        & 0.25        & 1 Hz               \\
                WIN\_SPD     & Wind Speed       & knots  & $V_w$        & 1.0        & 2 Hz               \\
                WIN\_DIR     & Wind Direction       & $^\circ$  & $\delta$        & 0.0039        & 2 Hz               \\
                FF1     & Fuel Flow of the Engine\#1       & lbs/h  & $f_1$        & 0.001        & 1 Hz               \\
                FF2     & Fuel Flow of the Engine\#2       & lbs/h  & $f_2$        & 0.001        & 1 Hz               \\
			   AOAL     & Angle of Attack Left        & $^\circ$   & $\alpha_l$ & 0.3516      & 4 Hz               \\
			   AOAR     & Angle of Attack Right       & $^\circ$   & $\alpha_r$ & 0.3516      & 4 Hz               \\ \hline
		\end{tabular}
	}
	\label{QARtable}
\end{table*}

\section{THE CONSTANT-GAIN EQUATION ERROR METHOD (CG-EEM) FRAMEWORK}

\begin{figure}[!htbp]
        \centering
    \includegraphics[width=0.6\linewidth]{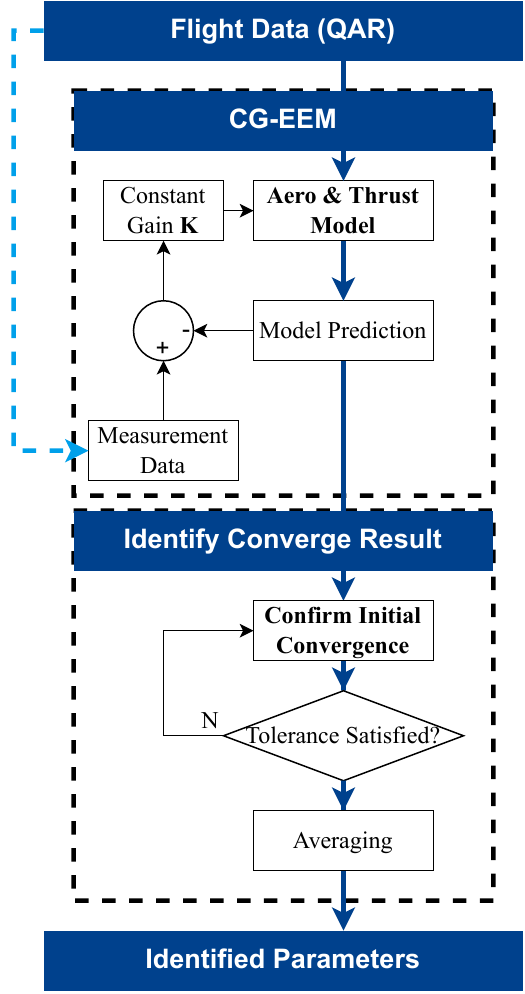}
    \caption{Flowchart of the proposed aerodynamic parameter identification framework, from raw QAR data to final identified aerodynamic parameters.}
    \label{fig:flow}
\end{figure}

This section details the framework developed to identify aerodynamic parameters from in-service QAR data. The methodology is named the Constant-Gain Equation-Error Method (CG-EEM) to reflect its two core design principles, which were chosen specifically to overcome the data limitations discussed in Section~\ref{QAR}. The overall architecture of the framework is depicted in Fig.~\ref{fig:flow} and consists of three main stages: Data Extraction, Parameter Estimation, and Convergence Analysis.

\subsection{The Decoupled Equation-Error Formulation}

The CG-EEM framework is built upon a decoupled, Equation-Error Method (EEM) formulation. It is a well-established principle in system identification. The EEM, as a form of least-squares, can yield biased parameter estimates when the input signals are corrupted. In this case, they are the measured states from QAR data ($\alpha$, $q$, $\theta$, $V$) corrupted by measurement noise.

Despite this potential for bias, the EEM was deliberately chosen over unbiased alternatives like the Output-Error Method (OEM). The fundamental barrier, as discussed in Section~\ref{introduction}, is that OEM requires propagating the aircraft state through a dynamic model, which is infeasible without accurate, time-varying moments of inertia—data not available in QAR. Therefore, we make a pragmatic trade-off: accepting a potentially small and consistent bias in exchange for a robust and tractable framework that bypasses the need for unavailable model parameters. This choice allows us to directly use the measured states as inputs to an algebraic model of the aircraft's forces. The estimation task is thus simplified to identifying the parameters of this force model that best explain the measured longitudinal acceleration $a_x$ and vertical acceleration $a_z$.

\subsubsection{State and Parameter Vectors}

The longitudinal motion of the aircraft is described by a state vector $\mathbf x$:
\begin{equation}
    \mathbf{x} = [\alpha, q, \theta, V]^T
\end{equation}

where $\alpha$ is the angle of attack, $q$ is the pitch rate, $\theta$ is the pitch angle, and $V$ is the true airspeed.

The aerodynamic model is defined by a vector of parameters $\mathbf\Theta$ to be estimated. Based on a standard polynomial representation of the aircraft's forces and moments, the parameter vector is defined as:
\begin{equation}
    \mathbf{\Theta} = [C_{L0}, C_{L\alpha}, C_{LM}, C_{D0}, C_{DL}, C_{TV}]^T
\end{equation}

This vector includes coefficients for lift, drag, and a term for calculating the thrust of the engine.

\subsubsection{Aerodynamic and Thrust Model}

Aerodynamic forces (lift $L$, drag $D$) are calculated at each time step. The model employs a linear structure for lift and a parabolic structure for drag, which is a standard and effective representation for conventional aircraft in the cruise phase of flight.

The Lift Coefficient $C_L$ is modeled as:
\begin{equation}\label{cl}
    C_L = C_{L0} + C_{L\alpha}\alpha + C_{LM}V_M
\end{equation}

The Drag Coefficient $C_D$ is modeled using the standard parabolic drag polar. This structure is chosen for its strong physical basis, representing drag as the sum of a baseline parasitic component $C_{D0}$ and a lift-induced component that varies with the square of the lift coefficient $C_L$:
\begin{equation}\label{polar}
    C_D = C_{D0} + C_{DL}{C_L}^2
\end{equation}

The aerodynamic model presented does not explicitly include control surface deflections (e.g., elevator, ailerons). This is a deliberate design choice to enhance robustness of the framework for two primary reasons. First, the focus of this study is on quasi-steady, wings-level cruise flight. During this phase, control inputs are minimal and primarily used for small trim adjustments, meaning their physical effect on the overall low-frequency aerodynamic forces is negligible.

Second, and equally important, the control surface signals recorded in operational QAR data during cruise often exhibit a very low signal-to-noise ratio. The small, high-frequency movements are frequently dominated by sensor noise and quantization effects. Including these noisy signals as inputs to the estimator could introduce more instability and variance than the information they provide, potentially degrading the accuracy of the primary parameter estimates. Therefore, by excluding these low-quality inputs, we simplify the model and simultaneously improve its robustness against the imperfections of real-world operational data.

It is important to note that alternative drag model formulations were also investigated during this research, including those with linear dependencies on the angle of attack \cite{EKFUKF}. However, our analysis consistently showed that these alternative structures were unable to provide stable convergence and resulted in physically inconsistent force estimations, particularly in the longitudinal channel. The parabolic polar, which correctly links drag to the generation of lift, proved to be essential for achieving a globally consistent solution across both the longitudinal and vertical axes. Therefore, this physically-grounded model was selected for our final framework.

Since direct thrust measurements are unavailable in QAR data, engine thrust $T$ is calculated from the measured fuel flow $f$ and the mach speed $V_M$. The core of the model is the Thrust Specific Fuel Consumption (TSFC), which relates fuel consumption to the thrust produced. While TSFC is a complex function of altitude, Mach number, and engine settings, research has shown that for the specific case of high-altitude, quasi-steady cruise, its dependency simplifies significantly. As concluded by \cite{TSFC}, we adopt a linear model for TSFC as a function of Mach speed, and therefore the thrust is calculated as:
\begin{equation}
    T = \frac{f}{TSFC}
\end{equation}
where $TSFC$ is calculated by 
\begin{equation}
    TSFC = T_0+C_{TV}V_M
\end{equation}
In this formulation, the measured fuel flow $f$ and Mach speed $V_M$ are inputs from the QAR data. The unknown thrust coefficient $C_{TV}$ is unknown and estimated simultaneously with the aerodynamic coefficients.

\subsubsection{Measurement Model}

The measurement vector $\mathbf z$ consists of the available sensor readings. The measurement model, $h(\mathbf x, \mathbf\Theta)$, predicts these sensor readings based on the current state and parameters.
\begin{equation}
    \mathbf{z}_k = h(\mathbf{x}_k, \mathbf{\Theta}_k) + \mathbf{v}_k
\end{equation}

where $\mathbf{v}_k$ is the measurement noise, assumed to be zero-mean Gaussian with covariance $\mathbf R$. The predicted measurement vector is:
\begin{equation}
    \hat{\mathbf{z}}_k = [\alpha, q, \theta, V, a_x, a_z]^T
\end{equation}

The longitudinal ($a_x$) and vertical ($a_z$) accelerations are calculated as:
\begin{align}
    a_x = \frac{-D \cos(\alpha) - L \sin(\alpha) + T \cos(\sigma)}{m}\label{ax}\\
    a_z = \frac{-D \sin(\alpha) + L \cos(\alpha) + T \sin(\sigma)}{m}\label{az}
\end{align}

where $m$ is the aircraft mass and $\sigma$ is the thrust line offset angle.

\subsection{The Constant-Gain Parameter Estimator}

The second core principle of the CG-EEM is the use of a constant-gain estimator. Standard Recursive Least Squares (RLS) was found to be unsuitable for this application because its time-varying gain diminishes too rapidly, leading to premature convergence on the low-excitation cruise data.

To address this, our estimator is designed to maintain a consistent, non-vanishing gain throughout the analysis window. While structured similarly to RLS, it differs in a crucial way: the parameter covariance matrix $\mathbf P$ is not recursively updated. Instead, the apriori covariance $\mathbf P'$ used for the gain calculation is reset at each time step $k$ based on the initial uncertainty matrix $\mathbf P_0$. The update equations are as follows:
\begin{enumerate}
    \item \textbf{Compute Measurement Residual (Error)}
    
    The error $\mathbf{e}_k$ is the difference between the actual measurement $\mathbf{z}_k$ and the measurement predicted using the parameters from the previous step, $\mathbf{\Theta}_{k-1}$.
    $$
    \mathbf{e}_k = \mathbf{z}_k - h(\mathbf{x}_k, \mathbf{\Theta}_{k-1})
    $$
    \textit{Note: For this algorithm, the state vector $\mathbf{x}_k$ is taken directly from the measured data at time $k$, not from a state propagation model.}

    \item \textbf{Compute Measurement Sensitivity Matrix (Jacobian)}
    
    The matrix $\mathbf{H}_k$ describes how sensitive the predicted measurement is to a small change in each parameter. It is the partial derivative of the measurement function with respect to the parameter vector.
    $$
    \mathbf{H}_k = \left. \frac{\partial h}{\partial \mathbf{\theta}} \right|_{\mathbf{x}_k, \mathbf{\theta}_{k-1}}
    $$
    The Jacobian is computed numerically at each step using finite differences due to the complexity of the model.

    \item \textbf{Compute Gain with Constant Covariance}
    
    The parameter covariance matrix is invariant during whole process. And a Kalman-like gain, $\mathbf{K}_k$, is computed to blend the new information from the measurement residual with the current parameter estimate.
    $$
    \mathbf P'=\mathbf P_0
    $$
    $$
    \mathbf{S}_k = \mathbf{H}_k \mathbf{P}' \mathbf{H}_k^\top + \mathbf{R}
    $$
    $$
    \mathbf{K}_k = \mathbf{P} '\mathbf{H}_k^\top \mathbf{S}_k^{-1}
    $$

    \item \textbf{Update Parameters}
    
    The parameters are then updated:
    $$
    \mathbf{\Theta}_k = \mathbf{\Theta}_{k-1} + \mathbf{K}_k \mathbf{e}_k
    $$
\end{enumerate}

\subsection{Implementation and Tuning}

The CG-EEM framework was implemented in the Julia programming language. The estimator requires tuning of two key matrices: the initial parameter covariance $\mathbf P_0$ and the measurement noise covariance $\mathbf R$. The initial parameter covariance, $\mathbf P_0$, represents the uncertainty in the initial guess for the parameter vector $\mathbf\Theta$. It was initialized as a diagonal matrix with large values (e.g. $10^2$) on the diagonal. This signifies high initial uncertainty, allowing the estimator to freely adapt to the information in the data without being biased by starting point.

The measurement noise covariance $\mathbf R$ is a diagonal matrix representing the expected variance of the measurement noise. A sensitivity analysis was conducted to determine the effect of $\mathbf R$ on the estimator's performance. It was found that the CG-EEM framework is remarkably robust to the choice of this parameter. For a typical initial parameter covariance $\mathbf P_0$ with diagonal elements on the order of 1e2, the final parameter estimates remained virtually unchanged for values of $\mathbf R$ ranging from 1e-3 to 1e1.

\subsection{Convergence Criteria and Final Parameter Extraction}\label{sec:conv}

The CG-EEM estimator produces a time-series of parameter estimates for each analyzed flight segment. While the constant-gain nature of the algorithm allows it to provide reasonably accurate real-time parameter tracking, the primary focus of this paper is on monitoring the average performance characteristics of aircraft over a stable flight segment, rather than its instantaneous dynamics. Therefore, to enable robust statistical analysis across a fleet, a systematic and objective procedure was established to extract a single, representative value for each parameter from the time-series output. This process explicitly addresses the question of how to ensure the estimator has truly converged to a stable and meaningful value.

The procedure involves two steps. First, an initial transient period is discarded, corresponding to the first 60\% of the time steps in a segment. Allowing the estimator ample time to move from its initial uninformed state towards a stable solution.

Second, convergence is objectively verified on the latter 40\% of the segment, defined as the "convergence window". Certainty of convergence is not assumed; it is statistically tested. For each estimated parameter, the Coefficient of Variation (CV) is calculated within this window. The CV, defined as the ratio of the standard deviation $\sigma$ to the absolute mean $|\mu|$, is a normalized measure of dispersion:
\begin{equation}
    CV=\dfrac{\sigma}{|\mu|}
\end{equation}

The Coefficient of Variation is chosen as the convergence criterion because it provides a normalized, dimensionless measure of dispersion, making it suitable for comparing the stability of different parameters with varying scales. A low CV value indicates that the parameter estimate has stabilized with low variance relative to its mean, signifying robust convergence.

An estimated parameter is considered to have successfully converged if its CV within the convergence window is less than a predefined threshold. The choice of this threshold accounts for the inherent identifiability of different physical effects. Based on the sensitivity and variance analysis presented in Section~\ref{sec:performance}, a stricter threshold of 1\% is used for the more readily identifiable lift-related parameters, while a threshold of 10\% is used for the drag-related parameters, whose estimation is more sensitive to measurement noise.

This objective criterion ensures that the final value is not derived from a drifting or excessively noisy estimate. Flights containing segments where any parameter fails to meet its respective CV threshold are flagged for manual inspection. It is important to clarify the interpretation of such instances, which are typically associated with unmodeled dynamics from significant turbulence or banking maneuvers.

In these dynamic, off-nominal conditions, the underlying assumption of slowly-varying aerodynamic coefficients, inherent to our linear performance model, may no longer hold. The CG-EEM framework, being an adaptive estimator, will attempt to track these rapid changes, causing the parameter estimates to fluctuate rather than settle. Consequently, they fail the CV-based convergence test not because the algorithm has failed, but because the concept of a single, stable "converged" parameter loses its physical meaning in such flight regimes.

Since the primary goal of this work is fleet-wide performance monitoring, which requires stable and comparable metrics, we deliberately focus on the quasi-steady cruise segments where aerodynamic characteristics are stable. This focus allows for the extraction of meaningful, consistent baseline parameters essential for long-term degradation analysis.

Once convergence is confirmed, the single representative value for each aerodynamic parameter is calculated as the mean of the estimates within this stable convergence window. This mean value is then used for all subsequent fleet-wide statistical analyses.

\section{PERFORMANCE AND SCALABILITY EVALUATION}

This section presents the validation and application of the CG-EEM framework. First, integrity and superiority of the algorithm over alternatives are verified using a simulated dataset with a simulated known ground truth. The validated framework is then applied to a large dataset of operational QAR flights to demonstrate its performance and utility.

\subsection{Verification on Simulated Data}

Before its application to operational data, the CG-EEM framework was subjected to a rigorous verification study using simulated data. The objective of this study was twofold: first, to confirm fundamental ability of the algorithm to identify known "ground truth" parameters from a realistic but controlled dataset, and second, to quantify the estimator's robustness to sensor noise.

\subsubsection{Simulation Setup}\label{sec:sim}

A realistic flight scenario was constructed to serve as the testbed. The state trajectory (airspeed, angle of attack, etc.) was taken directly from a 200-second quasi-steady cruise segment of a real Airbus A321 flight, thereby preserving the low-excitation characteristics of the target application.

The "ground truth" aerodynamic and thrust parameters for the simulation were pre-defined, with values chosen to be representative of the A321 fleet based on the aggregated results of this study as listed in Table \ref{tab:A321}. Using this state trajectory and the ground truth parameters, a "perfect" set of noise-free forces was generated. Finally, a pseudo-QAR dataset was created by adding artificial, zero-mean Gaussian noise to all state and force signals to mimic the noise characteristics of real sensor data. The intensity of this added noise was controlled to evaluate the estimator's performance, as described in Section~\ref{sec:noise}.

\subsubsection{Performance and Identifiability of CG-EEM}\label{sec:performance}

After applying the framework, the final identified parameters were used to predict the longitudinal and vertical accelerations.

\begin{figure}[!htbp]
    \centering
    \includegraphics[width=\linewidth]{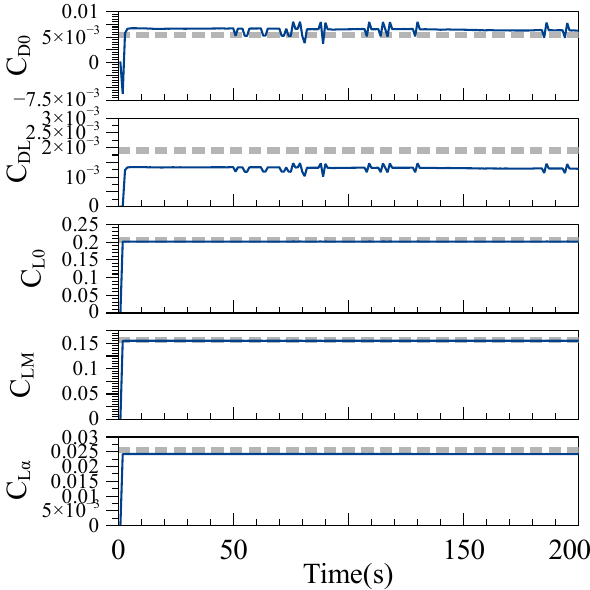}
    \caption{CG-EEM parameter convergence on simulated data. The time histories of the key aerodynamic parameter estimates (solid blue lines) are shown. The estimates demonstrate rapid and stable convergence from an initial 0 values to the known ground truth values (dashed grey lines). This verifies the fundamental integrity and accuracy of the CG-EEM algorithm.}
    \label{fig:simc}
\end{figure}

Fig.~\ref{fig:simc} shows the time history of the key parameter estimates. The algorithm demonstrates excellent performance, with all parameters converging rapidly and stably from their initial 0 value towards the known ground truth values.

Table~\ref{tab:sim} provides a quantitative comparison between the final identified parameter values (averaged over the stable convergence window) and the pre-defined true values. The identified lift and thrust-related parameters are recovered with high accuracy, showing minimal bias. The drag parameters $C_{D0}$ and $C_{DL}$ are also identified with reasonable accuracy, though they exhibit a slightly higher bias and a significantly larger Coefficient of Variation (CV). This difference in identifiability is expected: the drag forces are much smaller than the lift forces, making their corresponding parameters inherently more sensitive to measurement noise. This sensitivity is, in fact, a manifestation of the classic thrust-drag identifiability challenge, where small errors in the much larger thrust and lift forces can have a significant relative impact on the estimated drag. This result provides the quantitative justification for using a tiered CV threshold (e.g., 1\% for lift parameters and 10\% for drag parameters) in our convergence criteria as described in Section~\ref{sec:conv}.

\begin{table}[!htbp]
    \centering
    \caption{Verification of CG-EEM on Simulated Data}
    \label{tab:sim}
    \begin{tabular}{lcccc} % lcr for alignment
    \hline
    \textbf{Parameter} & \textbf{True} & \textbf{Identified} & \textbf{Std. Dev. (in window)} & \textbf{CV} \\
    \hline
    $C_{L0}$      & 0.2050 & 0.2016 & 1.83$\times10^{-5}$& 0.01\%\\
    $C_{L\alpha}$ & 0.0256 & 0.0242 & 0.15$\times10^{-5}$& 0.01\%\\
    $C_{LM}$      & 0.1570 & 0.1551 & 1.42$\times10^{-5}$& 0.01\%\\
    $C_{D0}$      & 0.0054 & 0.0063 & 4.23$\times10^{-4}$& 6.66\%\\
    $C_{DL}$      & 0.0019 & 0.0013 & 2.42$\times10^{-5}$& 3.42\%\\
    $C_{TV}$      & 0.0329 & 0.0285 & 2.50$\times10^{-4}$& 0.88\%\\
    \hline
    \end{tabular}
\end{table}

\begin{figure}[!htbp]
    \centering
    \includegraphics[width=\linewidth]{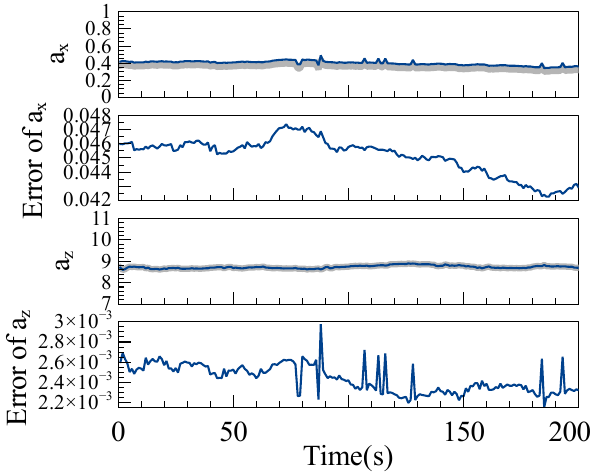}
    \caption{Fidelity of the converged model on simulated data. The top panels compare the accelerations predicted by the final identified model against the pseudo-QAR measurement data (grey). The bottom panels show the time history of the residual error. The model's predictions closely track the noisy measurements, and the residual errors are small, confirming that the identified parameter set provides a high-fidelity representation of the system dynamics.}
    \label{fig:PA}
\end{figure}

As a secondary verification, the fidelity of the final identified model was assessed by comparing its predicted accelerations against the pseudo QAR measurement data. Fig.~\ref{fig:PA} shows this comparison. The model's predictions (blue line) closely track the pseudo measurements (grey line), and the residual errors shown in the bottom panels are relatively small. The higher average error in the longitudinal acceleration $a_x$ compared to the vertical acceleration $a_z$ is consistent with the lower signal-to-noise ratio associated with the smaller drag forces. 

The mean residual for the longitudinal acceleration is 0.045 m/s$^2$, which is slightly larger than magnitude of vertical's. However, this magnitude must be contextualized by the sensor's physical limitations. The data acquisition system records longitudinal acceleration with a granularity of 0.0039g, which corresponds to a quantization step of 0.038 m/s$^2$.

The fact that the average residual error is on the same order of magnitude as the sensor's fundamental resolution is a critical finding. It indicates that the CG-EEM framework has successfully captured the underlying physical dynamics to the fullest extent possible, leaving a residual that is dominated by irreducible quantization noise rather than model inadequacy. This provides strong confidence in the fidelity of the identified aerodynamic parameters.

\subsubsection{Performance on Different Noise Level}\label{sec:noise}

To rigorously quantify the influence of sensor noise on the estimation accuracy and consistency of the CG-EEM, a Monte Carlo analysis was performed. The simulation described in Section~\ref{sec:sim} was repeated 100 times for several different noise levels. The "noise level" is defined as a multiplier applied to the standard deviation of the baseline artificial noise added to all signals.

The characteristics of this noise were determined empirically from the most stable, non-maneuvering segments of the real A321 QAR data. For each signal, the standard deviation of the real measured data was calculated in these quiet segments. This measured standard deviation was then used to define the baseline ($1\times$) for the artificial Gaussian random noise added in the simulation. This random noise was subsequently passed through a quantization process matching the known granularity of the corresponding QAR sensor. For example, the measured standard deviation for longitudinal acceleration was found to be 0.0026g (0.0255 m/s$^2$), which was set as the baseline for its random noise component.

Noise levels ranging from $0.5\times$ (half baseline noise) to $10.0\times$ were tested. For each of the 100 runs at a given noise level, a new, unique random noise sequence was generated.

\begin{figure}[!htbp]
    \centering
    \includegraphics[width=\linewidth]{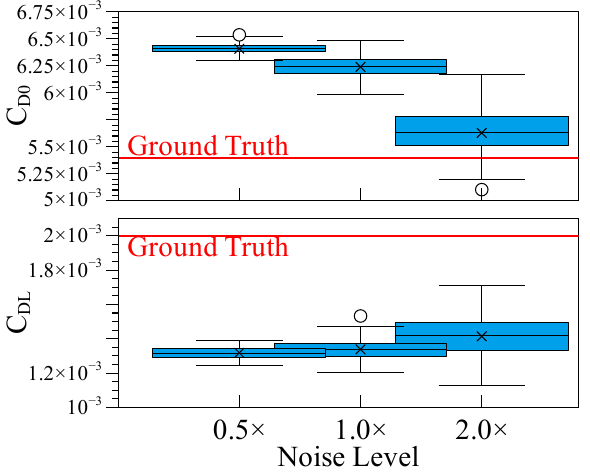}
    \caption{Distribution of Estimated Parameters at Varying Noise Levels. The box plots illustrate the distribution of the 100 estimated values for $C_{D0}$ and $C_{DL}$ at three different noise levels, compared against their known ground truth (red line). The increasing size of the distributions demonstrates the expected decrease in precision with higher noise.}
    \label{fig:noise}
\end{figure}

The aggregated results of this analysis are presented in Table~\ref{tab:noise} and Fig.~\ref{fig:noise}. Table~\ref{tab:noise} summarizes the key statistics for the estimated parameters, including the bias and the standard deviation of the estimates across the 100 runs. The analysis reveals a complex and important trade-off between estimation bias and variance.

\begin{table}[!htbp]
    \centering
    \caption{Monte Carlo Analysis of CG-EEM Performance vs. Noise Level (N=100 runs)}
    \label{tab:noise}
    \begin{tabular}{lcccc} % lcr for alignment
    \hline
    \textbf{Parameter} & \textbf{True} & \textbf{Noise Level} & \textbf{Mean} & \textbf{Std. Dev.} \\
    \hline
     & & $0.5\times$& 0.0063 & $4.533\times10^{-5}$\\
     & & $1.0\times$& 0.0062 & $9.437\times10^{-5}$\\
     $C_{D0}$&0.0054& $2.0\times$& 0.0056 & $2.072\times10^{-4}$\\
     & & $5.0\times$& 0.0020 & $6.513\times10^{-4}$\\
     & & $10.0\times$& -0.0046 & $1.068\times10^{-3}$\\
    \hline
     & & $0.5\times$& 0.0013 & $3.171\times10^{-5}$\\
     & & $1.0\times$& 0.0013 & $5.791\times10^{-5}$\\
     $C_{DL}$&0.0019& $2.0\times$& 0.0014 & $1.274\times10^{-4}$\\
     & & $5.0\times$& 0.0020 & $4.972\times10^{-4}$\\
     & & $10.0\times$& 0.0053 & $2.059\times10^{-3}$\\
    \hline
     & & $0.5\times$& 0.2016 & $7.180\times10^{-5}$\\
     & & $1.0\times$& 0.2016 & $1.399\times10^{-4}$\\
     $C_{L0}$&0.2050& $2.0\times$& 0.2021 & $2.516\times10^{-4}$\\
     & & $5.0\times$& 0.2036 & $6.465\times10^{-4}$\\
     & & $10.0\times$& 0.2050 & $1.431\times10^{-3}$\\
    \hline
    \end{tabular}
\end{table}

As expected, the estimation variance (the spread of the estimates, visualized by the size of the box plots) increases monotonically with the noise level. This confirms that higher noise consistently leads to lower precision.

However, the estimation accuracy (the proximity of the mean estimate to the true value) follows a more nuanced trend. At very low noise levels ($0.5\times$), a small but consistent bias is observed. As the noise level is increased to $1.0\times$ and $2.0\times$, this bias is systematically reduced, with the mean of the estimates moving closer to the ground truth. We conclude this is a dithering effect, where the random noise mitigates the systematic bias introduced by signal quantization.

At very high noise levels ($5.0\times$ and $10.0\times$), this dithering benefit becomes overshadowed by the large random component. While the mean estimate may remain close to the true value, the variance becomes so large that the reliability of any single estimation is significantly degraded.

This demonstrates that there is an optimal range of noise for this estimation problem. The inherent noise level of the real QAR data, modeled here as the 1.0x baseline, appears to fall within this beneficial range. It is noisy enough to mitigate the worst effects of quantization bias, but not so noisy as to make the parameter identification problem intractable. This dual finding is a powerful demonstration of the framework's robustness. It not only handles random noise gracefully, but its accuracy is actively enhanced by it in the presence of the signal quantization found in all real-world QAR data. This analysis provides strong confidence that the CG-EEM is exceptionally well-suited for its target application and that the results obtained from real operational data are reliable.

\subsection{Application to Operational Flight Data}

Having rigorously validated the performance of CG-EEM framework on simulated data, we now apply it to a large operational dataset to demonstrate its real-world performance. The analysis comprises over 200 commercial flights and is presented in three parts. First, the framework's fidelity is examined in detail on a representative A321 flight segment. Second, a fleet-wide statistical analysis is conducted on 135 flights from 33 unique Airbus A321 aircraft to assess the consistency and distribution of the identified parameters. Finally, the framework's generality and scalability are demonstrated through its application across five distinct airliner types: the Airbus A320, Airbus A321, Boeing 737, Boeing 777, and Boeing 787.

\subsubsection{Case Study on Single Flight}

Having verified the CG-EEM framework and justified its design through simulation, we now demonstrate its performance on real operational data. A 500-second segment of quasi-steady, wings-level cruise from an Airbus A321 flight was selected as a representative case study.

\begin{figure}[!h]
        \centering
    \includegraphics[width=\linewidth]{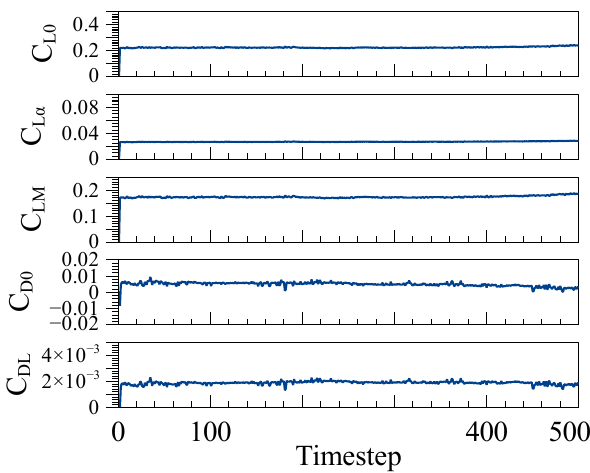}
    \caption{Time histories of key aerodynamic parameter estimates. All parameters demonstrate rapid convergence to stable values.}
    \label{fig:params}
\end{figure}

Fig.~\ref{fig:params} presents the time histories of the six key aerodynamic and thrust parameters estimated by the CG-EEM framework for this flight. All parameters exhibit excellent convergence behavior, reaching stable, non-drifting values well within the analysis window. This confirms that the constant-gain approach is effective at extracting a consistent solution from the noisy, low-excitation QAR data.

\begin{figure}[!h]
        \centering
    \includegraphics[width=\linewidth]{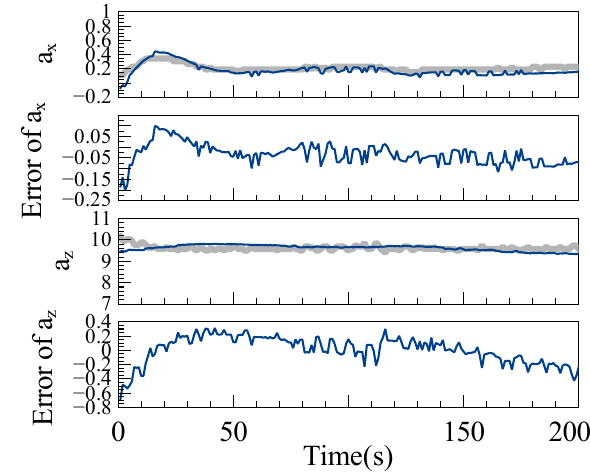}
    \caption{Model Fidelity and Residual Error of the case study flight segment.}
    \label{fig:states}
\end{figure}

The fidelity of the final converged model was then assessed by comparing its predicted accelerations against the actual measurements recorded in the QAR data. As shown in Fig.~\ref{fig:states}, the predictions of the model for both longitudinal acceleration $a_x$ and vertical acceleration $a_z$ closely track the real-world measurements.

The combination of stable parameter convergence and high model fidelity on this representative flight demonstrates that the CG-EEM framework can successfully and reliably identify an aircraft's aerodynamic characteristics from standard operational data.

\subsubsection{Fleet-Wide Statistical Analysis}

The robustness and consistency of the framework were validated on an extensive dataset comprising 135 flights from 33 unique Airbus A321 aircraft. The converged parameter estimates from each flight were aggregated for statistical analysis.

A detailed statistical summary is provided in Table~\ref{tab:A321}, where the mean values are shown to be physically plausible and align with expected values for this aircraft class.

\begin{table}[h!]
    \centering
    \caption{Statistical Summary of Identified Parameters for the A321 Fleet (135 flights)}
    \label{tab:A321}
    \begin{tabular}{l c c c c}
    \hline
    \textbf{Parameter} & \textbf{Mean} & \textbf{Std. Dev.} & \textbf{Max} & \textbf{Min} \\
    \hline
    $C_{L0}$      & 0.2050  & 0.0374 & 0.2830  & 0.1502  \\
    $C_{L\alpha}$ & 0.0256 & 0.0060 & 0.0391 & 0.0156 \\
    $C_{LM}$      & 0.1570  & 0.0294 & 0.2231  & 0.1161  \\
    \hline
    $C_{D0}$      & 0.0054 & 0.0076 & 0.0169 & 0.0014 \\
    $C_{DL}$      & 0.0019 & 0.0011 & 0.0055 & 0.0008 \\
    \hline
    $C_{TV}$      & 0.0329 & 0.0042 & 0.0440 & 0.0251 \\
    \hline
    \end{tabular}
\end{table}

\begin{figure}[!h]
    \centering
    \includegraphics[width=0.8\linewidth]{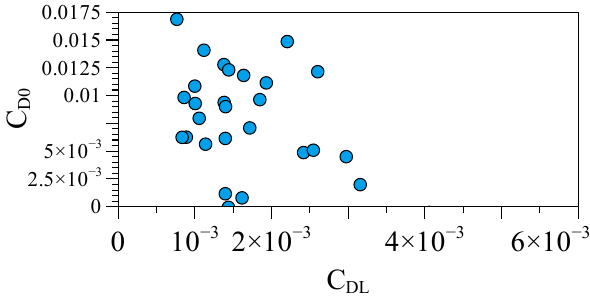}
    \caption{Correlation Analysis of Zero-Lift coefficient $C_{D0}$ and Induced Drag Coefficient $C_{DL}$.}
    \label{fig:A321}
\end{figure}

To further assess the model's fidelity, the correlation between the zero-lift and induced drag coefficients was examined. The scatter plot in Fig.~\ref{fig:A321} shows no significant correlation between the identified values of $C_{D0}$ and $C_{DL}$. This statistical independence is a critical result, as it indicates the estimator is successfully distinguishing between the two main sources of drag and not simply compensating for model errors by adjusting both parameters. This confirms the structural integrity of the underlying aerodynamic model.

This comprehensive analysis validates the framework's potential as a reliable, low-cost tool for monitoring fleet-wide aerodynamic performance. It is critical to emphasize that the consistency and statistical independence shown here are not merely artifacts of the estimator. A fundamental challenge in this domain is the thrust-drag ambiguity, where multiple incorrect parameter sets can yield statistically plausible results. We have addressed this challenge directly in our follow-on research by developing a novel validation criterion based on the dynamic response of the model's acceleration bias. This subsequent work rigorously confirms that the physically plausible solutions identified by the CG-EEM framework are indeed unique and physically correct, rather than being one of many mathematically ambiguous possibilities. This verification provides the ultimate confidence in using these identified parameters for real-world engineering and operational applications.

\subsubsection{Framework Generality Across Multiple Aircraft Types}

To validate the generality of the framework, it was applied without modification to a dataset spanning five aircraft types: the Airbus A320/A321, Boeing 737, Boeing 777, and Boeing 787. The aggregated results, shown in Table~\ref{tab:fleet}, confirm the robustness and physical consistency.

\begin{table*}[h!]
    \centering
    \caption{Fleet-wide statistics for identified drag parameters across five aircraft types.}
    \label{tab:fleet}
    \begin{tabular}{l c c c c c}
    \hline
    \textbf{Aircraft Type} & \textbf{Flights} & \textbf{Mean $C_{D0}$} & \textbf{Std. Dev. $C_{D0}$} & \textbf{Mean $C_{DL}$} & \textbf{Std. Dev. $C_{DL}$} \\
    \hline
    Airbus A320   &  28 & 0.0149 & 0.0032 & 0.0009 & 0.0004 \\
    Airbus A321   & 135 & 0.0054 & 0.0076 & 0.0019 & 0.0011 \\
    Boeing 737    &  10 & 0.0287 & 0.0006 & 0.0011 & 0.0001 \\
    \hline
    Boeing 777    &  28 & 0.0361 & 0.0123 & 0.0026 & 0.0103 \\
    Boeing 787    &  78 & 0.0119 & 0.0017 & 0.0003 & 0.0001 \\
    \hline
    \end{tabular}
\end{table*}

As shown in Table~\ref{tab:fleet}, the standard deviation for the identified parameters within each fleet is small relative to the mean. This low variance demonstrates that the estimation framework is robust, providing consistent results across numerous flights of the same aircraft type.

The framework is also sensitive enough to distinguish between aircraft types based on their aerodynamic characteristics. The mean values for $C_{D0}$ and $C_{DL}$ are distinct for each fleet and align with physical expectations. For instance, the mean zero-lift drag ($C_{D0}$) is higher for the wide-body B777 than for the narrow-body aircraft. Furthermore, the results reflect technological differences: the composite-built, modern B787 has a significantly lower mean $C_{D0}$ than the older B777, highlighting its superior aerodynamic efficiency. This confirms the model’s ability to capture meaningful physical differences between aircraft.

\section{CONCLUSION}

This paper has demonstrated a practical and effective framework, the Constant-Gain Equation-Error Method (CG-EEM), for the challenging task of aerodynamic parameter estimation from operational QAR data. The core contribution of this work lies not in proposing a more complex algorithm, but in the insight that a robust solution emerges from the careful and synergistic pairing of a physically-grounded aerodynamic model with an estimator specifically designed for the unique characteristics of the data. By systematically confronting the real-world constraints inherent in QAR data---most notably the absence of critical mass property parameters like moments of inertia and the pervasively low-excitation nature of cruise flight---the CG-EEM framework successfully transforms these vast and noisy datasets from a latent liability into a valuable, actionable asset for detailed engineering analysis.

More profoundly, this tailored approach achieves more than just algorithmic stability. As confirmed by our subsequent validation studies, the specific structure of the CG-EEM, combined with the underlying physics it models, effectively navigates the classic problem of thrust-drag ambiguity. This allows the framework to converge not just to a mathematically plausible solution, but to the unique, physically correct parameter set, a critical step that has long been a barrier to confident model validation without ground-truth force measurements.

The implications of this work, therefore, extend far beyond the academic exercise of parameter identification. The demonstrated ability to reliably, affordably, and most importantly, verifiably monitor the true aerodynamic health of an entire fleet represents a significant operational advantage for airlines. It provides a solid, data-driven foundation for a new generation of proactive maintenance strategies, allows for the refinement of fuel efficiency initiatives with aircraft-specific performance data, and enhances operational safety by enabling the early detection of performance degradation.

Future research will build upon this robust foundation. The immediate path forward involves incorporating higher-fidelity, nonlinear thrust models to capture a wider range of engine behaviors. Ultimately, this research establishes a validated, low-cost, and scalable pathway toward continuous, fleet-wide aerodynamic performance monitoring. By providing a proven mechanism to ensure the physical fidelity and uniqueness of its results, this work offers a powerful new tool for developing truly self-validating flight dynamic models, paving the way for more intelligent and autonomous aviation systems.

\section*{Acknowledgments}
This work was supported by National Key Research and Development Program of China [grant number 2023YFB4302903]; and National Nature Science Foundation of China [grant number U52572363].

\bibliographystyle{ieeetr} 
\bibliography{APE}

\newpage

\section*{Biography Section}
\begin{IEEEbiography}[{\includegraphics[width=1in,height=1.25in,clip,keepaspectratio]{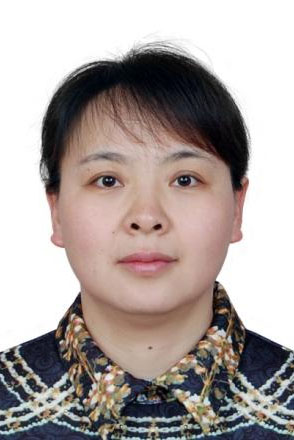}}]{Ruiying Wen}{\space}received the B.S., M.S. and Ph.D degree from Beihang University(Beijing University of Aeronautics and Astronautics), Beijing, China.

She is an associate professor and master supervisor of College of Air Traffic Management, Civil Aviation University of China. She has long been engaged in the research of aircraft aerodynamics and performance.
\end{IEEEbiography}%

\begin{IEEEbiography}[{\includegraphics[width=1in,height=1.25in,clip,keepaspectratio]{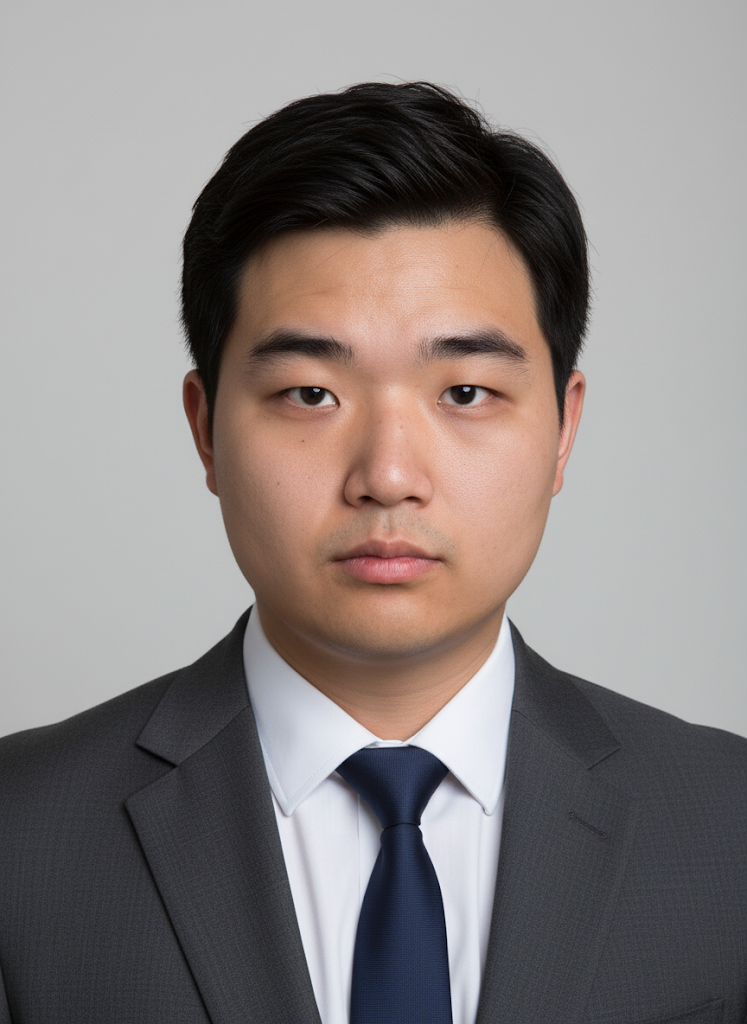}}]{Yuntao Dai}{\space}received the B.S degree in College of Transportation from Northeast Forestry University, Harbin, China in 2022. He is currently pursuing the M.S degree with College of Air Traffic Management from Civil Aviation University of China.

His research interests currently focus on aircraft aerodynamics, performance and flight simulation.
\end{IEEEbiography}

\begin{IEEEbiography}[{\includegraphics[width=1in,height=1.25in,clip,keepaspectratio]{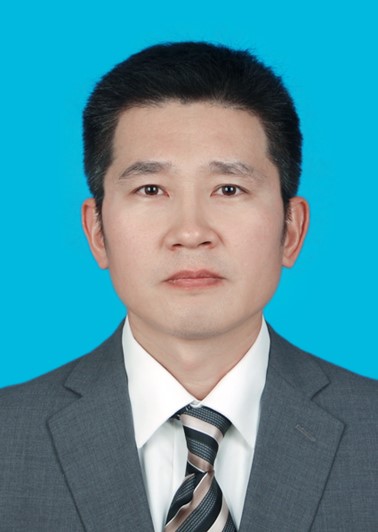}}]{Hongyong Wang}{\space}received the Ph.D degree in transportation engineering from Nanjing University of Aeronautics and Astronautics, Nanjing, China.

He is a professor and master supervisor of College of Air Traffic Management, Civil Aviation University of China. He has long been working in the field of aircraft performance, complex network and airspace management.
\end{IEEEbiography}

\vfill

\end{document}